\documentclass[aps,prl,twocolumn,superscriptaddress,showpacs]{revtex4-1}

\usepackage{graphicx}
\usepackage{dcolumn}
\usepackage{bm}
\usepackage{float}
\usepackage{amssymb}
\usepackage{amsmath}
\usepackage{mathtools}
\usepackage{lipsum}
\usepackage{soul}
\usepackage{upgreek}

\DeclarePairedDelimiter\ket{\lvert}{\rangle}
\usepackage{hyperref}
\usepackage[mathlines]{lineno}

\usepackage{notes2bib}

\bibnotesetup{  
note-name = ,
use-sort-key = false
}

\usepackage[dvipsnames]{xcolor}

\begin{document}

\title{Low Energy Excitations of a 1D Fermi Gas with Attractive Interactions}

\author{Aashish Kafle}
    \affiliation{Department of Physics and Astronomy, Rice University, Houston, Texas 77005, USA}
    
\author{Ruwan Senaratne}
    \affiliation{Department of Physics and Astronomy, Rice University, Houston, Texas 77005, USA}

\author{Danyel Cavazos-Cavazos}
\affiliation{Department of Physics and Astronomy, Rice University, Houston, Texas 77005, USA}

\author{Hai-Ying Cui}
\affiliation{Innovation Academy for Precision Measurement Science and Technology, Chinese Academy of Sciences, Wuhan 430071, China}
\affiliation{University of Chinese Academy of Sciences, Beijing 100049, China}

\author{Thierry Giamarchi}
\affiliation{Department of Quantum Matter Physics, University of Geneva, Geneva, Switzerland}

\author{Han Pu}
\affiliation{Department of Physics and Astronomy, Rice University, Houston, Texas 77005, USA}

\author{Xi-Wen Guan$^\dagger$}
\affiliation{Innovation Academy for Precision Measurement Science and Technology, Chinese Academy of Sciences, Wuhan 430071, China}
\affiliation{Department of Fundamental and Theoretical Physics, Research School of Physics, Australian National University, Canberra ACT 0200, Australia}

\author{Randall G. Hulet$^\dagger$}
\affiliation{Department of Physics and Astronomy, Rice University, Houston, Texas 77005, USA}
\email{randy@rice.edu}

\date{\today}
\begin{abstract}

The low-energy excitations of a two-component repulsive Fermi gas confined to one dimension are linear dispersing spin- and charge-density waves whose respective propagation velocities depend on the strength and sign of their interaction. Quasi-1D fermions with attractive interaction realize the Luther-Emery liquid, which exhibits a rich array of phenomena, many of which are qualitatively different from those exhibited by their repulsive counterpart \cite{Giamarchi_Book}. We use a Feshbach resonance to access attractive interactions with $^6$Li atoms. We measured the spin and charge dynamic structure factors using Bragg spectroscopy and find that, contrary to repulsive interactions, the spin wave propagates faster than the charge density wave, thus producing an inversion of the classic spin-charge separation. We also find that a small spin polarization strongly suppresses the spin gap in the measured Bragg spectra.  Evidence for pairing are a
reduction in spin correlations with increasing attraction and RF spectra consistent with an atom/molecule mixture.

\end{abstract}
\maketitle

The Tomonaga-Luttinger liquid (TLL) theory \cite{Tomonaga1950,Luttinger1963,Haldane1981,Voit1993,Giamarchi_Book,lieb1965exact,cazalilla2011one} is a powerful framework for the study of quantum many-body systems in one dimension (1D). In particular, it can be used to characterize the correlations and excitation spectra of 1D repulsive spin-1/2 fermions. At zero temperature $T$, the small-momentum excitations are collective charge- and spin-density waves (CDW and SDW, respectively) that are linearly dispersing but with propagation velocities that depend on the interaction strength. The fact that the SDW travels more slowly than the CDW results in a spin-charge separation of the collective interactions, which has been experimentally studied in quasi-1D solid-state materials with momentum-resolved tunneling measurements \cite{Auslaender2002, Auslaender2005, Jompol2009} and angle-resolved photoemission spectroscopy \cite{Segovia1999,Kim1996,Kim2006}. More recently, ultracold atomic fermions have been used in investigations of spin-charge separation \cite{Hilker2017,Ernie2018,salomon2019,Vijayan2020,Senaratne2022}, where the ability to precisely control and quantify the system parameters imparts new insight and illustrates the abilities of quantum simulation with ultracold atoms. In Ref.~\citenum{Senaratne2022}, we used Bragg spectroscopy to independently measure the dynamic structure factors of the low-energy SDW and CDW modes as a function of the strength of a repulsive interaction \bibnote{Although there is no ``charge’’ degree of freedom, but rather an atomic density, we adopt the charge label to be consistent with common usage}. The effects of nonlinearities are small \cite{Imambekov2012}, but were nonetheless observed by comparison of the data to precise theory \cite{Senaratne2022,Cavazos2023}. 

\begin{figure}[t!]
\includegraphics[width=0.48\textwidth]{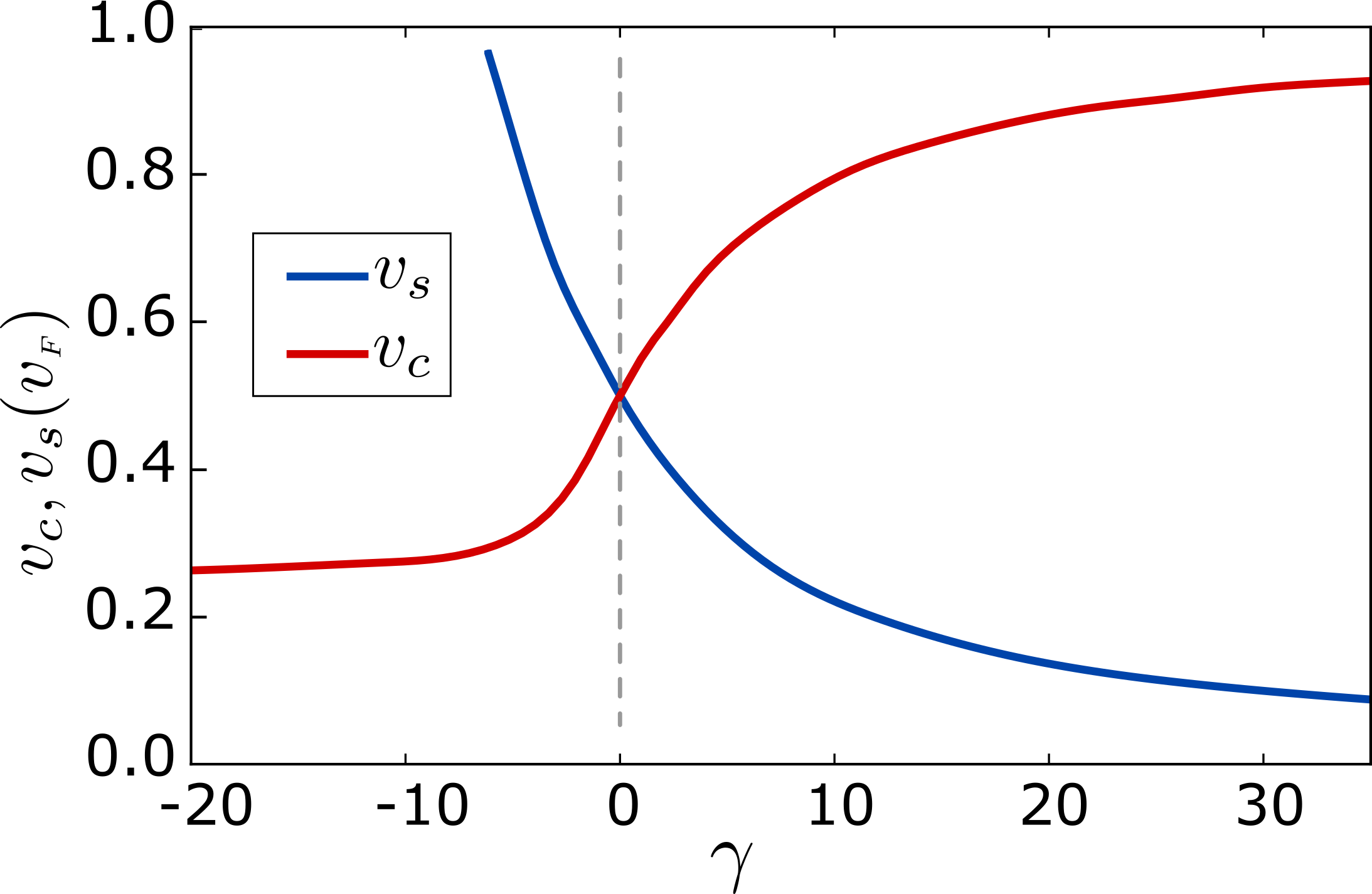}

\caption{Spin and charge velocities ($v_s$ and $v_c$) vs.\@ the dimensionless interaction strength $\gamma$ for the spin-1/2 1D Fermi gas. The solid lines are the
velocities obtained by numerically solving the Bethe-ansatz equations. The classic spin-charge separation is inverted in the attractive regime ($\gamma < 0$), where pairing is expected to cause $v_c$ to plateau at half the noninteracting value and the spin mode to disappear resulting from the onset of a gap. Figure adapted from Ref.~\citenum{Batchelor2006}.}
\label{fig:BA}
\end{figure}

More attention has been given to the case of repulsive rather than attractive interactions, perhaps because electron-electron interactions are usually repulsive in electronic materials. The attractive case, known as a Luther-Emery liquid \cite{Luther1974}, that exhibits a gapped spin dispersion $\omega(p) = \sqrt{\Delta^2 + v_s^2p^2}$ in the absence of a magnetic field, where $\Delta$ is the energy gap and $v_s$ the spin velocity, presents a richness of phenomena in its own right. Figure \ref{fig:BA} shows the result of a Bethe ansatz calculation of the speed of sound for both spin-waves (blue) and charge density waves (red) \cite{Guan2013}, as a function of $\gamma$, the dimensionless interaction parameter for a quasi-1D gas \cite{Dunjko2001}. These calculations indicate that for attractive interactions ($\gamma < 0$), the spin-mode velocity will be larger than that of the charge-mode, an inversion of the velocity hierarchy observed in the repulsive case. Strong attraction results in tightly bound pairs with mass twice the fermion mass, and thus the charge velocity $v_c \approx \frac{1}{4}v_F(1+\frac{1}{|\gamma|})$ for $\gamma \ll -1$ approaches half of its non-interacting value \cite{Guan2013}. Here $v_F = \hbar \pi \rho_\mathrm{1D}/m$ is the Fermi velocity with the total 1D linear density $\rho_\mathrm{1D}$ and atomic mass $m$. The spin-mode velocity $v_s \approx \sqrt{\Delta} (1+ \frac{2}{|\gamma|})$ is expected to diverge with increasing attraction as a gap opens in the spin sector \cite{Giamarchi_Book, Guan2013}.


Evidence for a Luther-Emery phase was previously indicated by the transport conductance in a quantum wire with a filling of two fermions per site\cite{Lebrat2018}, where the bound pairs of fermions form a band insulator with the band gap less than the superfluid gap. In our work, we probe the low-lying spin and charge excitations for low to moderately strong attractive interactions using Bragg spectroscopy and confirm that the spin-mode propagation velocity is indeed larger than that of the charge-mode in accordance with spin-gapped excitations of the  Luther-Emery liquid. In addition, we find evidence for an atom/molecule mixture using RF spectroscopy at finite temperature. 

Our experimental methods have been discussed in several prior publications \cite{Randy2020, Senaratne2022}. Briefly, we prepare a nominally spin-balanced mixture of $^6$Li atoms in the energetically lowest ($\ket{1}{}$) and third-lowest ($\ket{3}{}$) hyperfine sublevels in an isotropic optical trap at a temperature $T\approx 0.1\,T_F$, where $T_F$  is the Fermi temperature. Subsequently, we load the atoms into a 2D optical lattice with a depth of $15\,E_r$, where $E_r = 1.43\,\mu$K is the recoil energy of a lattice photon of wavelength 1.064$\,\mu$m. The resulting 2D array of quasi-1D tubes has transverse and axial confinement frequencies of $\omega_{\perp}/2\pi \simeq 227.5$ kHz and $\omega_{z}/2\pi \simeq 1.34$ kHz, respectively. A typical ensemble contains a total of $6.5\times10^4$ atoms, with the most probable and maximum tube occupancies of $\sim$35 and $\sim$70 atoms, respectively.

The number distribution across this array is non-uniform due to the Gaussian curvature of the optical beams and, in general, it also depends on the interaction strength. We compensate for the differences in the number distribution for different interaction strengths by applying a focused repulsive green (532 nm) laser beam along each of the lattice directions during ramp-up \cite{Mathy2012,Hart15}. By adjusting the intensity of these anti-confining beams, we can adjust the overall confinement potential of the lattice and thus minimize the variation of the number distribution across the array with interaction. We measure this distribution using \textit{in situ} phase-contrast imaging \cite{Randy2020} and then perform an inverse Abel transform to extract the radial number profile.

The $s$-wave scattering length $a_s$ is tunable by proximity to a broad Feshbach resonance at 690 G \cite{Zurn13}. Due to a large negative background scattering length, there is a zero-crossing in the scattering length on the low-field, Bardeen-Cooper-Schrieffer (BCS) side of the resonance at 568 G. Although the minimum value of $a_s$ below the Feshbach resonance is -890$\,a_0$, where $a_0$ is the Bohr radius, we are limited to no less than -500$\,a_0$ by the bandwidth of the Bragg laser frequency lock. This technical limitation also applies to the strongly attractive regime above the Feshbach resonance. A more fundamental limitation, imposed by three-body recombination, is $|a_s| \lesssim 600\,a_0$.

Bragg spectroscopy is a well-established method for measuring the dynamic structure factors (DSF's), $S(q,\omega)$, in ultracold atomic gases. Several experiments have used Bragg spectroscopy with fermions in 1D \cite{Senaratne2022,Ernie2018,Pagano2014} and in 3D \cite{Veeravalli2008,Hoinka2012}. In our implementation, the hyperfine sublevels $\ket{1}{}$ and $\ket{3}{}$ of the $2S_{1/2}$ ground state of the $^6$Li atom constitute the (pseudo) spin-up and spin-down states.  A two-photon coherent Bragg scattering process imparts momentum $\hbar q$ and energy $\hbar \omega$ while leaving the internal atomic states unchanged. The angle between the two Bragg laser beams determines $q$, which is fixed to be $q \simeq 0.2 k_F$, a small fraction of the Fermi momentum $k_F$, to ensure that the scattering process is well into the linear dispersion regime. The energy is varied by changing the relative frequency difference $\omega$ between the two Bragg beams.

Spin or charge modes may be separately excited by the appropriate choice of the detuning of the two Bragg beams. The charge mode is excited with a symmetrical detuning, for which the two beams are detuned by a frequency that is large compared to the frequency between the $\ket{1}{}$ and $\ket{3}{}$ sublevels, while if the two beams are detuned equally, but opposite in sign between the $\ket{1}{}$ and $\ket{3}{}$ states, they can excite a spin mode \cite{Hoinka2012,Senaratne2022}. Spontaneous scattering is minimized by making the detunings as large as possible compared to the natural linewidth. For the charge mode, we use the 2S-2P transition at 671 nm with a large 10 GHz detuning that renders spontaneous scattering irrelevant. The antisymmetric detuning needed to create spin excitations, on the other hand, is too small to avoid spontaneous scattering on the usual 671 nm transition. We mitigate this problem by switching to the $2S_{1/2}-3P_{3/2}$ transition at 323 nm \cite{Duarte2011}, which has an 8 times narrower natural linewidth than the principal transition at 671 nm \cite{Senaratne2022}. Additionally, the frequency separation between $\ket{1}{}$ and $\ket{3}{}$ is approximately twice that as for $\ket{1}{}$ and $\ket{2}{}$. We have demonstrated less than 10\% atom loss from spontaneous scattering, and we have previously shown that there is no adverse effect on the shape of the measured Bragg spectrum \cite{Senaratne2022}. We adjust the angles to ensure that $\lvert \vec{q}\rvert = 1.47\,\mu\rm{m}^{-1}$ for either excitation wavelength.

\begin{figure}[t!]
\includegraphics[width=0.48\textwidth]{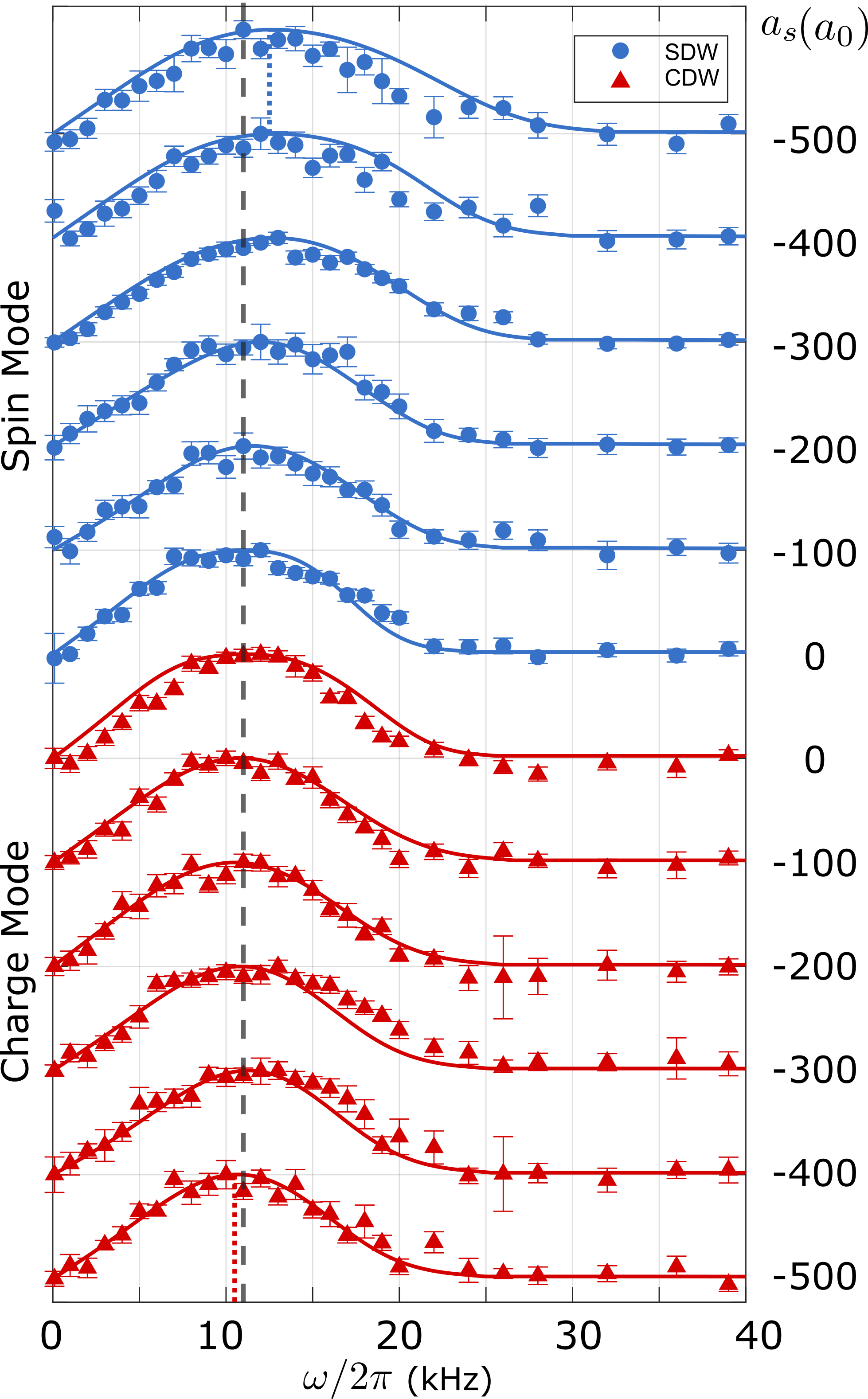}
\caption{Normalized Bragg spectra corresponding to $S_c(q,\omega)$ (red triangles) and $S_s(q,\omega)$ (blue circles). Each data point is the average of at least 20 separate experimental shots. Error bars represent the standard error obtained by bootstrapping \cite{Efron1979}. Vertical lines show the extracted peak frequency $\omega_p$ for the non-interacting case (dashed black) and $\omega_p$ for $a_s = -500$ $a_0$ in the case of spin (dotted blue line) and charge (dotted red line). The solid red (blue) lines are fits to the calculated charge-mode (spin-mode) spectra from the Bethe ansatz solution, with fitting parameters $T = 250$ nK and an overall scaling.}
\label{fig:spectra}
\end{figure}


The Bragg beams are applied for 200 $\mu$s, and their intensity is chosen to ensure that the momentum transfer is in the linear response regime. We then switch off the confinement and allow the atoms to expand freely for 150 $\mu$s, after which the cohort of atoms that received the Bragg kick is visibly displaced from the center of the atom cloud. We obtain the normalized Bragg signal by subtracting a background image, without a Bragg pulse, from a corresponding signal image, with a Bragg pulse.

\begin{figure}[h!]
\includegraphics[width=0.48\textwidth]{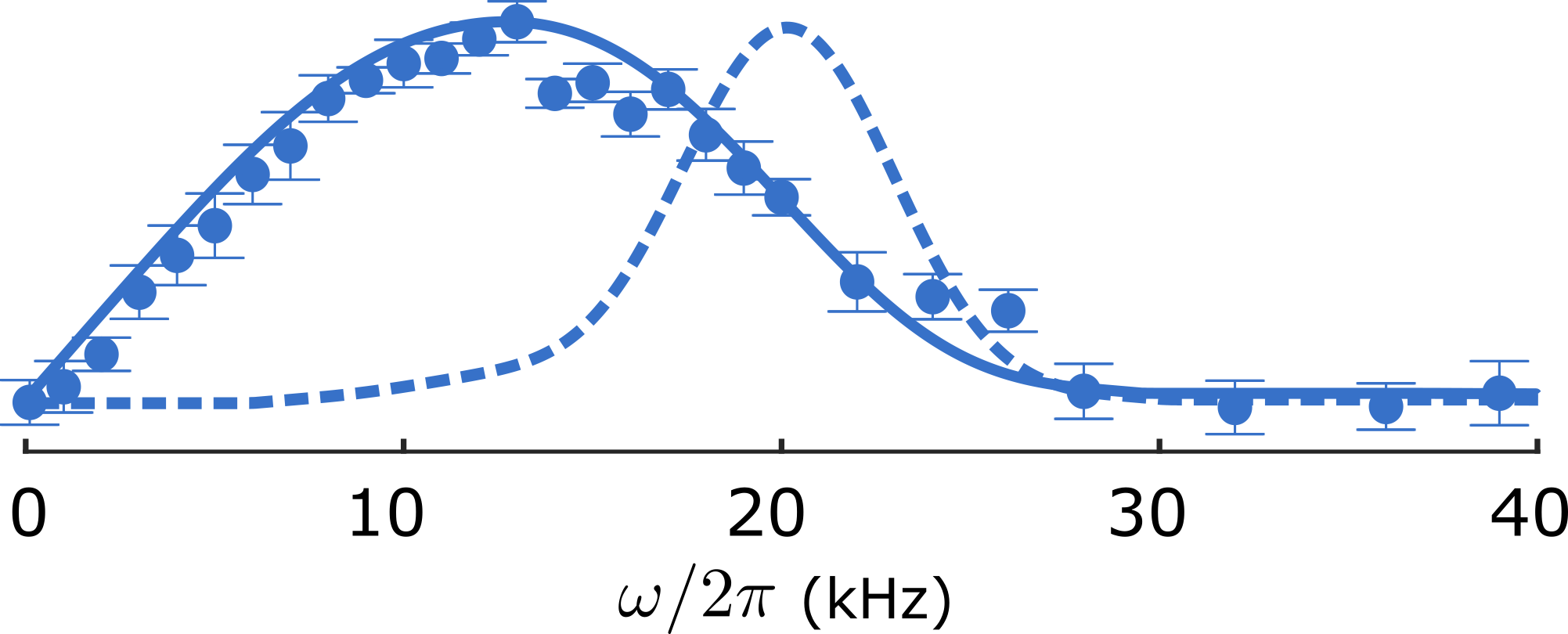}
\caption{Spin Bragg spectrum for $a_s=-300a_0$. The symbols are data and the solid line is the calculated spectrum by assuming a small polarization of $p=0.1$. These are reproduced from Fig.~\ref{fig:spectra}. The dashed line is the calculated spectrum assuming perfect spin balance with $p=0$. This clearly demonstrates the significant effects on the spin DSF by small spin imbalance.}
\label{fig:balance}
\end{figure}

Figure \ref{fig:spectra} shows the measured and calculated (see Methods \cite{supp}) Bragg spectra for both modes for $a_s$ ranging from -500 to 0 $a_0$. We estimate the DSF’s, as in our previous work \cite{Ernie2018,Senaratne2022}, by modifying the free-fermion DSF with the interaction-dependent Fermi velocity $v_F$ for a homogeneous density at $T=$ 0. Then, the charge- and spin-mode velocities $v_{c,s}$ are found as functions of $\gamma$ and $v_F$ using the Bethe ansatz. The interaction parameter $\gamma = {m g_1 (a_s)}/{\hbar^2 \rho_\mathrm{1D}}$, where $m$ is the atomic mass, $g_1(a_s)$ is the coupling strength of the quasi-1D pseudopotential \cite{Olshanii1998} and $\rho_\mathrm{1D}$ is the 1D density. We invoke the local density approximation (LDA) by summing contributions to the Bragg signal (DSF) from segments along each tube, and then by summing contributions from each tube in the ensemble. We account for the spectral broadening arising from the finite Bragg pulse duration. The fitting parameters are the independent peak-height normalizations, and a global temperature that is found to be $250$ nK. 

\begin{figure}[h]
\includegraphics[width=0.48\textwidth]{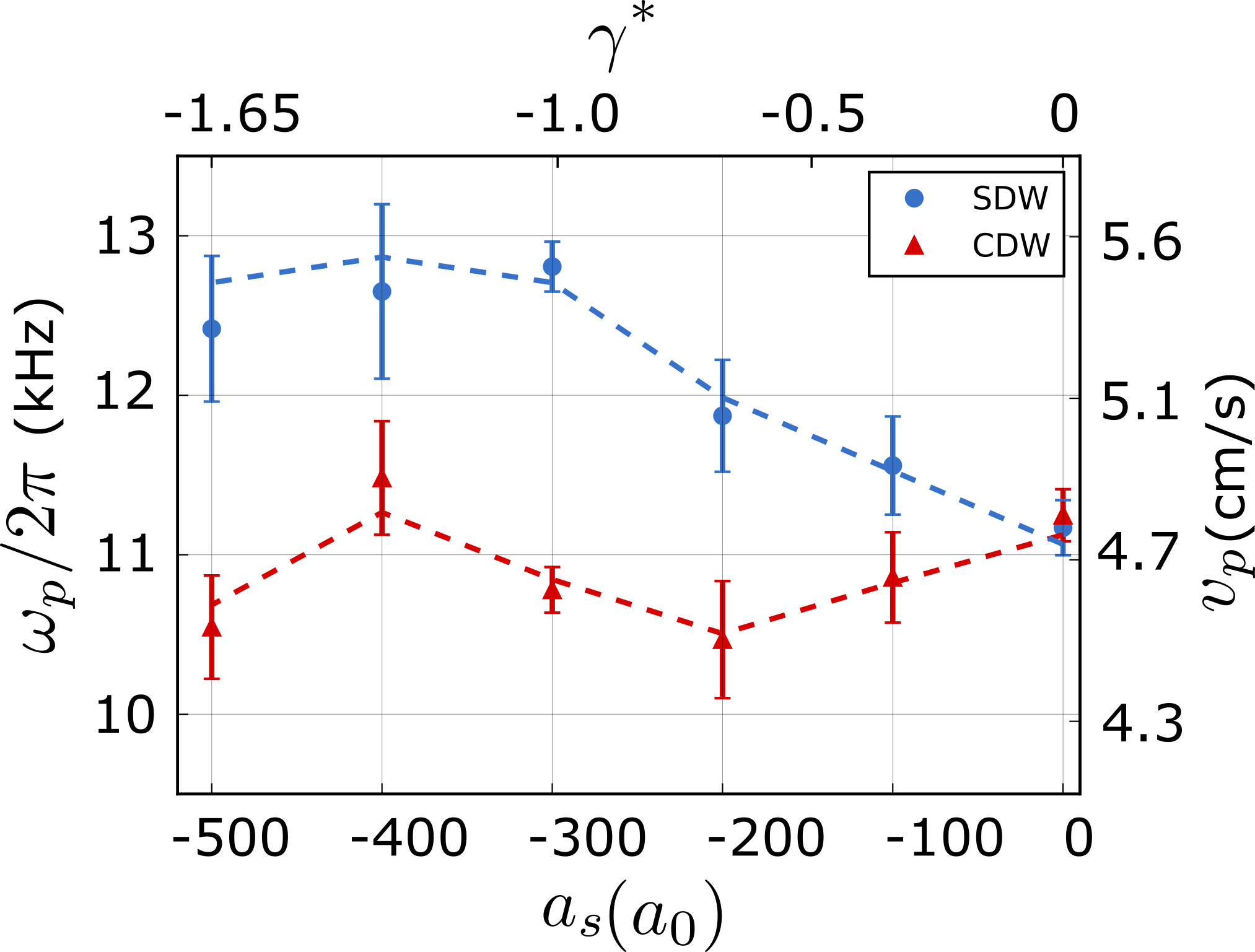}

\caption{Locations of frequencies, $\omega_p$, corresponding to the peak of the DSF's for charge (red triangles) and spin (blue circles) excitations extracted from Fig.\@ \ref{fig:spectra}. $\omega_p$ values were determined through fits of a parabolic function to the data points above 50\% of the maximum measured value, and error bars are statistical standard errors of the fit parameters. The corresponding speed of sound, $v_p=\omega_p/q$, is given by the right axis. The upper horizontal axis gives the interaction strength in terms of the interaction parameter $\gamma^*$, obtained from $\gamma$ evaluated at the center of a tube with an average occupancy of 35 atoms. The dashed lines show the calculated values of $\omega_p$ for the spin and charge modes from the Bethe ansatz solution. The point to point variations of $\omega_p$ arise from variations of the density profile for different interactions, see the Methods \cite{supp}.}
\label{fig:peaks}
\end{figure}


Our system consists, in general, of both paired and unpaired atoms. The latter originates from thermal excitations due to finite temperature and/or from the spin imbalance due to shot noise when preparing the sample. The spin DSF is dominated by the unpaired atoms, which fundamentally alters the excitation spectrum, thus connecting the spin excitations from gapped to gapless. This explains why no direct evidence of a spin gap can be seen in the spin spectrum in Fig.~\ref{fig:spectra}. The charge excitation, on the other hand, is always gapless regardless of whether the atoms are paired or not. For the temperature of our system ($T=250$ nK), we expect that the unpaired atoms are predominantly contributed by a spin imbalance with strong attraction ($|a_s|>300a_0$). In our theoretical calculation of the DSF, we have assumed a 10\% spin imbalance that is consistent with the expected shot noise. Such small imbalance has nearly no effect on the charge spectrum, but dramatically affects the spin spectrum with strong attraction. This is illustrated in Fig.~\ref{fig:balance} where we compare the calculated spin spectrum for $a_s=-300a_0$, with and without spin imbalance. The DSF without spin imbalance shows a spin gap and does not fit the data at all. For weaker attractive interaction ($|a_s|< 300a_0$), by contrast, small spin imbalance is not as important as thermal fluctuations. Our observations reveal that the atomic behavior of collective motion is induced by a few quasiparticles near the Fermi points, very similar to the way that the two-spinon continuum spectrum essentially determines the low-energy excitations in the long wavelength limit.  More details of our theoretical calculation can be found in Methods\cite{supp}.

The frequencies, $w_p$, at which each DSF is maximized are obtained from Fig.\@ \ref{fig:spectra} and plotted in Fig.\@ \ref{fig:peaks} vs.\@ $a_s$. Each value of $\omega_p$ corresponds to the most probable value of the mode velocity $v_\mathrm{p} = \omega_\mathrm{p}/q$ in the ensemble. As the magnitude of the interaction increases, the spin-mode velocity increases, in accordance with the exact Bethe ansatz results (dashed lines), and in agreement with the prediction for constant density (Fig.\@ \ref{fig:BA}). The charge-mode behavior is complicated by the increase in $\rho_\mathrm{1D}$ with increasing attraction. This factor contributes to the charge-mode velocity even for a tube with a fixed number of atoms, since $v_c$ tends to increase with increasing density. In contrast, the spin mode depends on unpaired fermions at low temperature due to the asymmetric detuning.  These are well modeled by our approximation of the charge-mode and spin spectra, as shown by the dashed lines in Fig.\@ \ref{fig:peaks}, see Methods.

The $v_{c,s}$ at 0 $a_0$ are equal to each other, as they should, but $\sim$10\% higher than reported for repulsive interactions \cite{Senaratne2022}. This difference arises from differences in the way we chose to use green light to compensate for the density inhomogeneity. The highest compensation intensity was used for the least repulsive interaction on the $\gamma>0$ side \cite{Senaratne2022}, while on the attractive side ($\gamma<0$), the greatest compensation corresponds to the highest attraction to make the density profiles for all scattering lengths similar. This causes a $\sim$20\% difference in density at $\gamma^* = 0$, and since $v_F \propto k_F \propto \sqrt{\rho_\mathrm{1D}}$, the result is a $\sim$10\% increase in $v_F$, as observed.

\begin{figure}[h]
\includegraphics[width=0.48\textwidth]{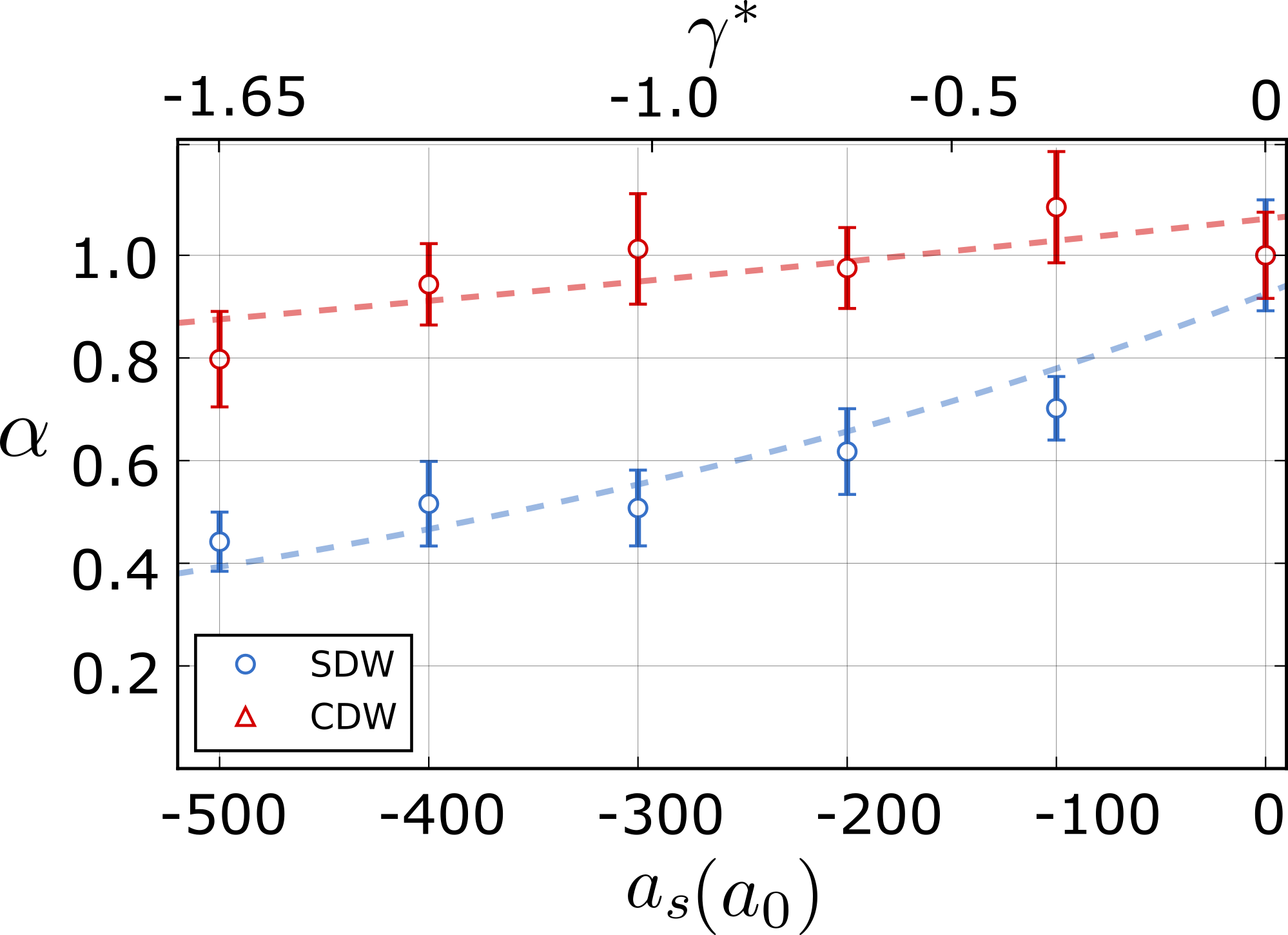}

\caption{Normalized peak height, $\alpha$, of the DSF obtained from a skew-Gaussian fit to the measured DSFs at different interactions. The error bars are obtained from statistical standard errors of the relevant fit parameters. The data points are normalized to $\alpha = 1$ for the measured DSF height at $0$ $a_0$. The dashed blue (red) line is an exponential fit in $\gamma^*$ to the spin (charge) mode.  }
\label{fig:peaksComparisionv2}
\end{figure}


 In Fig.\@ \ref{fig:peaksComparisionv2}, we show the normalized peak height, $\alpha$, obtained from the measured Bragg spectra. The spin mode shows a significant reduction in the correlation strength for $\gamma \lesssim -1$ compared to the charge mode. For repulsive interactions \cite{Senaratne2022}, the reduction of the measured peak height for either mode was small, even for $\gamma^* \simeq 4$. The decrease in spin correlations with increasing attraction signals the growth of the spin gap and the pair population predicted by the Luther-Emery model.


The Luther-Emery liquid is an atom-molecule mixture with the molecular binding energy, $\epsilon_b$, given by $\frac{a_{s}}{a_{\perp}} = \frac{\sqrt{2}}{\zeta(1/2, -\epsilon_b/2\hbar \omega_{\perp})}$, where $\zeta$ is the Hurwitz zeta function \cite{bergeman2003atom}, is small for interactions probed in this work, where $\epsilon_b \approx$ 12 kHz at -500 $a_0$. 
Here, a combination of factors, such as finite temperature, spin imbalance due to shot noise in sample preparation, as well as non-ideal adiabatic loading of the sample to 1D, results in a mixture of bound molecules and unbound atoms.
We employ RF spectroscopy to investigate the molecule fraction, with details shown in Methods. These results are consistent with the Bragg spectra. 


In summary, the observation of a higher spin-mode than charge-mode velocity confirms the inversion of the spin-charge separations for attractive interactions as compared with the repulsive case of interacting spin-1/2 1D fermions. This work shows the pair-breaking nature of the 1D attractive Fermi gas~\cite{orso2007attractive,liao2010spin,Batchelor2006,Guan2013,kim2018pair}. A small polarization substantially suppresses the spin-gap excitation as evident from the decreasing magnitude of the spin DSF with increasing attraction. This is a significantly different scenario from the two-spinon excitations that characterize the 1D repulsive Fermi gas. In order to observe the spin gap directly in the Bragg spectrum, we conclude that one needs to make a nearly perfect spin balance.




\bibliography{MainText-FINAL}

\begin{thebibliography}{41}
\bibitem [{1()}]{1}%
  \BibitemOpen
  \href@noop {} {}\bibinfo {note} {Although there is no ``charge’’ degree of freedom, but rather an atomic density, we adopt the charge label to be consistent with common usage}\BibitemShut {NoStop}%
\end{thebibliography}

\vspace{5mm} 

\begin{acknowledgments}
The authors dedicate this work to the memory of Professor Chen Ning Yang. R.G.H. thanks the NSF (Grant No. 2309362) for supporting this work. D. C.-C. acknowledges financial support from CONACyT (Mexico, Scholarship No. 472271). H.P. is supported by the NSF (Grant No. PHY-2513089) and the Welch Foundation (Grant No. C-1669).
X.W.G. and H.Y. C. acknowledge support from the NSFC key grants No. 92365202, No. 12134015, and the National Key R\&D Program of China under Grant No. 2022YFA1404102. X.W.G. is also partially supported by the Innovation Program for Quantum Science and Technology 2021ZD0302000. TG acknowledges support from the Swiss National Science Foundation under Division II (Grant No. 200020219400).

\end{acknowledgments}

\newpage

\end{document}


\title{Methods: Low Energy Excitations of a 1D Fermi Gas with Attractive Interactions 
     }
    \author{Aashish Kafle, Ruwan Senaratne, Danyel Cavazos-Cavazos, Hai-Ying Cui, Thierry Giamarchi, Han Pu, Xi-Wen Guan, and Randall G. Hulet}

	\maketitle
    \section{I. RF Spectroscopy}
\begin{figure}[h!]
\includegraphics[width=0.50\textwidth]{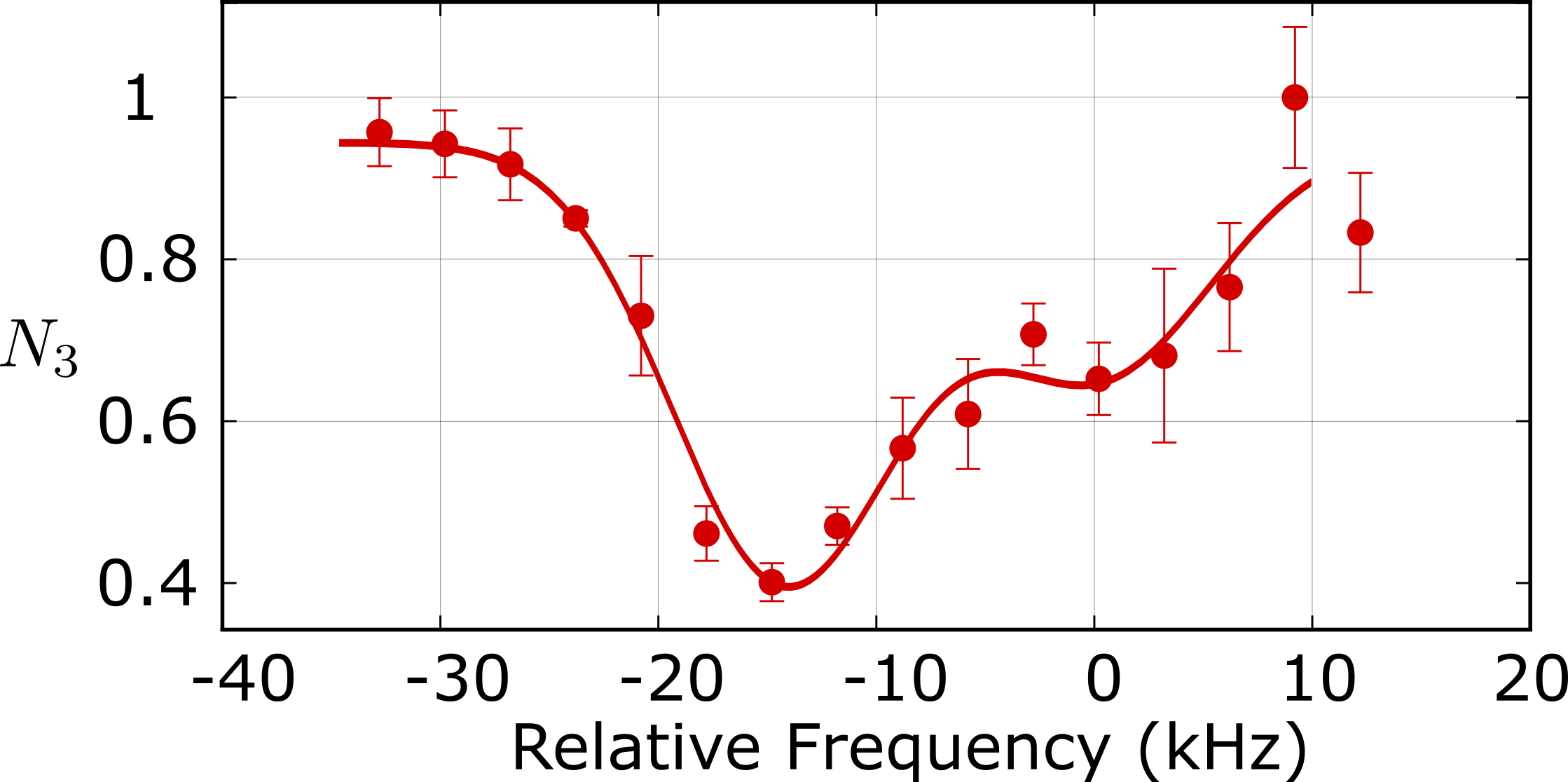}
\caption{Radio-frequency scan of a 1D sample prepared at $-500$ $a_0$ for a $\ket{1}{}$-$\ket{3}{}$ mixture. A 60$\,\mu$s RF pulse drives atoms in the energetically higher state $\ket{3}{}$ to a lower state $\ket{2}{}$, followed by state selective absorption imaging of the remaining population in $\ket{3}{}$. At this field of 510 G, the $a_{13}=$ -500 $a_0$, and the $\ket{1}{}$-$\ket{3}{}$ binding energy is $\sim$$12$ kHz. The atomic resonance is at $\delta \nu = 0$. The solid line is a double Gaussian fit to the measured RF spectra for a 15 $E_r$ 2D lattice.}
\label{fig:500a0_RFscan_doublegauss}
\end{figure}

   In order to probe molecules, we drive atoms in state $\ket{3}{}$ to the unpopulated state $\ket{2}{}$ and measure the remaining number in state $\ket{3}{}$. The spectrum at -500 $a_0$ is shown in Fig.\@ \ref{fig:500a0_RFscan_doublegauss}, where we see a dip in the state $\ket{3}{}$ population at zero detuning, which can be explained by the presence of free atoms. The dip at the lower detuning is near the calculated value of $\epsilon_b$. It matches the measured RF spectrum well.

	\section{II. Bethe ansatz solutions and local density approximation}
	
	We first examine the implications of inhomogeneous density distributions in a 1D trapped spin-1/2 Fermi gas in terms of Bethe ansatz equations. Consider a system of fermions with mass $m$ confined in a 1D harmonic potential with $V(x)=\frac{1}{2}m\omega^{2}_xx^{2}$. 
According to the local density approximation, we define the local chemical potential as 
	\begin{eqnarray}
		\mu\left(x\right)=\mu_0-V(x)\label{1}
	\end{eqnarray}
	where $ \mu_0$ is the chemical potential at the center of the trap. 
Therefore, the density profile of fermions in the quasi-1D tube is also nonuniform. 
The total number of fermions and the polarization in the tube are given by
	\begin{eqnarray}\label{2}
		\begin{aligned}
			N&=\int dx\ n(x), \\
			p&=(N_{\uparrow}-N_{\downarrow})/N=2\int dx \, m^z(x) /N,
		\end{aligned}
	\end{eqnarray}
respectively,
where $n(x)=2n_b(x) +n_u(x) $ is the total density, $n_u(x)$ ($n_b(x)$) is the density of unpaired (paired or bound) fermions, $m^z(x)$ is the magnetization density.
We make two comments here: (1) At zero temperature, all unpaired fermions are fully polarized, i.e. the polarization $p=\left(\int dx\ n_u(x)\right)/N$.
At finite temperature, this relation does not hold rigorously but is still a good approximation for sufficiently low temperatures; 
(2) The unpaired fermions show a ferromagnetic ordering for attractive interaction, as can be clearly seen from the thermal Bethe ansatz (TBA) equations. 
	
	For a homogeneous 1D system of interacting fermions, we can calculate the thermodynamic physical quantities exactly using the TBA equations. At zero temperature, the number densities of bound pairs and unpaired fermions are given by \cite{Guan:2007}
 \begin{eqnarray}\label{3}
	 	\begin{aligned}
	 		n_b&= \int_{-B}^B \sigma(\Lambda) d \Lambda, \\
	 		n_u&=\int_{-Q}^Q \rho(k) d k\\
	 		n&=2n_b+n_u=2 \int_{-B}^B \sigma(\Lambda) d \Lambda+\int_{-Q}^Q \rho(k) d k, \\
	 	\end{aligned}
	 \end{eqnarray}  
	 	where, for a given interaction strength $c$, 
        \begin{eqnarray}\label{6}
		\begin{aligned}
			\rho(k) & =\frac{1}{2 \pi}-\frac{1}{2 \pi} \int_{-B}^B \frac{|c| \sigma(\Lambda)}{c^2 / 4+(k-\Lambda)^2} d \Lambda, \\
			\sigma(\Lambda) & =\frac{1}{\pi}-\frac{1}{2 \pi} \int_{-B}^B \frac{2|c| \sigma\left(\Lambda^{\prime}\right)}{c^2+\left(\Lambda-\Lambda^{\prime}\right)^2 d \Lambda^{\prime}}-\frac{1}{2 \pi} \int_{-Q}^Q \frac{|c| \rho(k)}{c^2 / 4+(\Lambda-k)^2} d k,
		\end{aligned}
	\end{eqnarray}
        are distribution functions of the quasi-momentum of unpaired (paired) fermions, respectively; $Q$ ($B$) are the corresponding Fermi points determined by $\varepsilon_u\left(Q\right)=0=\varepsilon_b\left(B\right)$, with $\varepsilon_u$ and $\varepsilon_b$ being the dressed energies that are given by 
	\begin{eqnarray}\label{8}
		\begin{aligned}
			&\varepsilon_u(k)=k^2-\mu-\frac{1}{2}H-\int_{-B}^B a_1\left(k-\Lambda\right) \varepsilon_b\left(\Lambda\right) d \Lambda,\\
			 &\varepsilon_b(\Lambda)=2\left(\Lambda^2-\mu-\frac{1}{4} c^2\right)-\int_{-B}^B a_2\left(\Lambda-\Lambda^{\prime}\right) \varepsilon_b\left(\Lambda^{\prime}\right) d \Lambda^{\prime}-\int_{-Q}^Q a_1(\Lambda-k) \varepsilon_u(k) d k,
		\end{aligned}
	\end{eqnarray}
    where $H$ represents the magnetic field, and $a_\ell (x) \equiv \frac{1}{2\pi}\frac{\ell c}{\left(\ell c/2\right)^2 x^2}$. 
%

	 At low finite temperature $T$, we can define the particle densities through pressure: $n_{b(u)} = \partial p_{b(u)}/\partial \mu$, where
     \begin{eqnarray}\label{14}
		\begin{aligned}
			p_b&=\frac{k_BT}{\pi} \int_{-\infty}^{\infty} \ln \left(1+e^{-\varepsilon_b(\Lambda) / k_BT}\right) d \Lambda, \\
			p_u&=\frac{k_BT}{2 \pi} \int_{-\infty}^{\infty} \ln \left(1+e^{-\varepsilon_u(k) / k_BT}\right) d k, 
		\end{aligned}
	\end{eqnarray}
and the finite-temperature dressed energies are given by
	\begin{eqnarray*}
			\varepsilon_b(\Lambda)&=&2\left(\frac{\hbar^{2}}{2m}\Lambda^2-\mu-\frac{\hbar^{2}}{8m} c^2\right)+k_BT a_2 * \ln \left(1+e^{-\frac{\varepsilon_b(\Lambda)}{k_BT}} \right)+k_BT a_1 * \ln \left(1+e^{-\frac{\varepsilon_u(\Lambda) }{ k_BT} }\right), \\
			\varepsilon_u(k)&=& \frac{\hbar^{2}}{2m}k^2-\mu-\frac{1}{2}g\mu_B H+k_BT a_1 * \ln \left(1+e^{-\frac{ \varepsilon_b(k) }{ k_BT} }\right)-k_BT \sum_{m=1}^{\infty} a_m * \ln \left(1+e^{ -\frac{\varepsilon_m(k) }{ k_BT} }\right) \\
			\varepsilon_n(\lambda)&=& ng\mu_BH+{ k_BTa_n * \ln \left(1+e^{- \frac{\varepsilon_u(\lambda) }{ k_BT}}\right)}+  k_BT\sum_{m=1}^{\infty} T_{n m} * \ln \left(1+e^{-\frac{\varepsilon_m(\lambda) }{k_BT}}\right),
	\end{eqnarray*}
where the last equation represents the dressed energy of strings of length-$n$, showing a ferromagnetic ordering of a Heisenberg spin chain. This can be seen from the sign of the second term in the expression of $\varepsilon_n(\lambda)$ that coincides with the dressed energy of a ferromagnetic spin chain.  
Such a ferromagnetic ordering couples to the dressed energy of unpaired fermions through the last term in the equation for $\varepsilon_u(k)$.

	 We observe that for a homogeneous system, the particle number density and the polarization are determined by the chemical potential $\mu$ and the magnetic field $H$.
	For an inhomogeneous 1D tube,
we divide the tube into many small cells, each of which can be regarded as a local homogeneous subsystem characterized by the local chemical potential $\mu(x)$ and magnetization density $m^z(x) $. At low temperatures, the total number $N$ and polarization $p$ in the tube can therefore be calculated as
 \begin{eqnarray}\label{16}
	 	\begin{aligned}
	 		N&=\int dx\ n(x,\mu_0,H), \\
	 		p&=2\left(\int dx\ m^z(x,\mu_0,H)\right)/N\approx \left(\int dx\ n_u(x,\mu_0,H)\right)/N.  \label{polarization-LDA}
	 	\end{aligned}
	 \end{eqnarray}
In the experiment, a 2D array of tubes is realized, and we need to average over all the tubes.

\begin{figure}[H]
		\centering
		\subfigure[]{\includegraphics[scale=0.68]{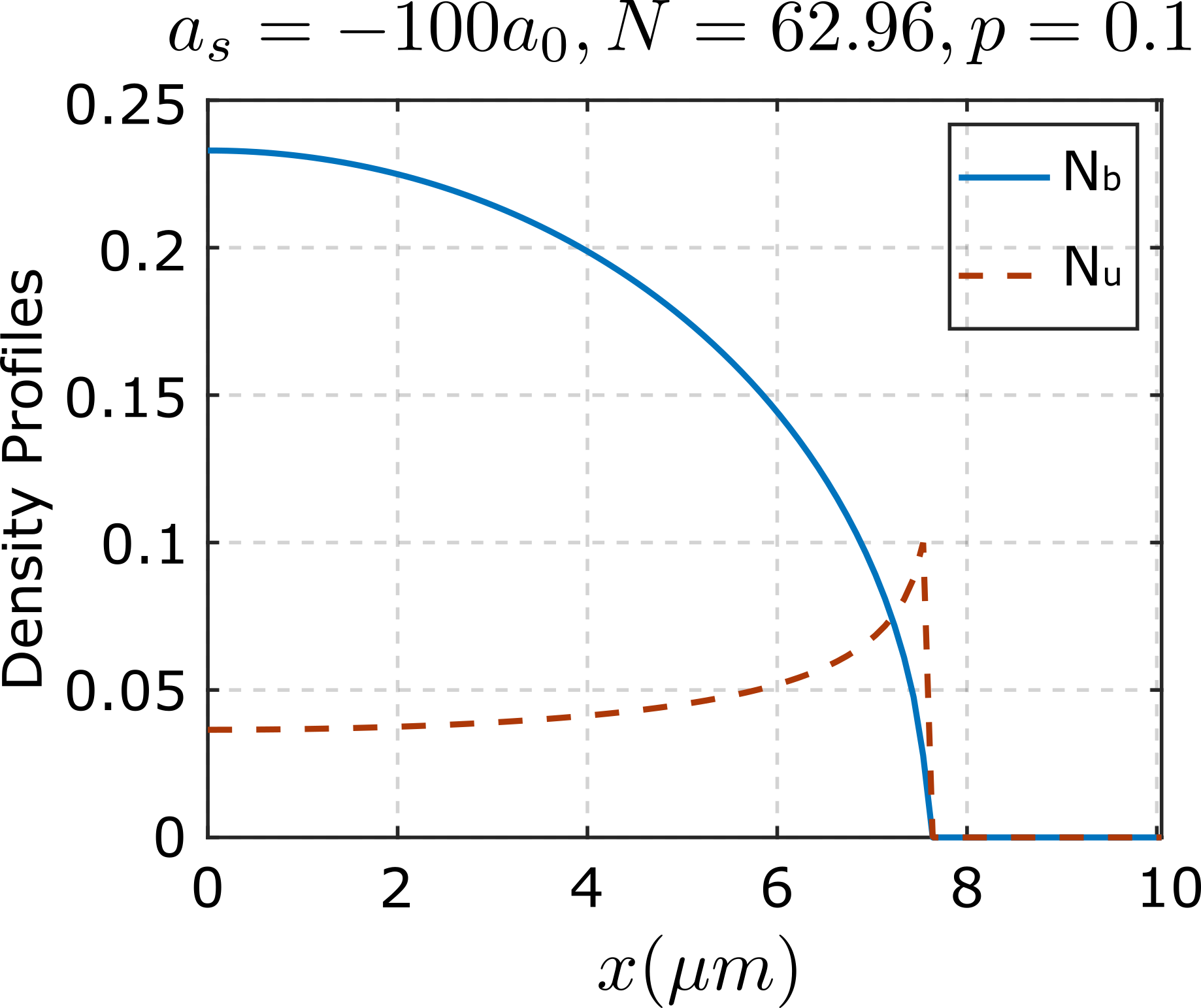}}
		\subfigure[]{\includegraphics[scale=0.68]{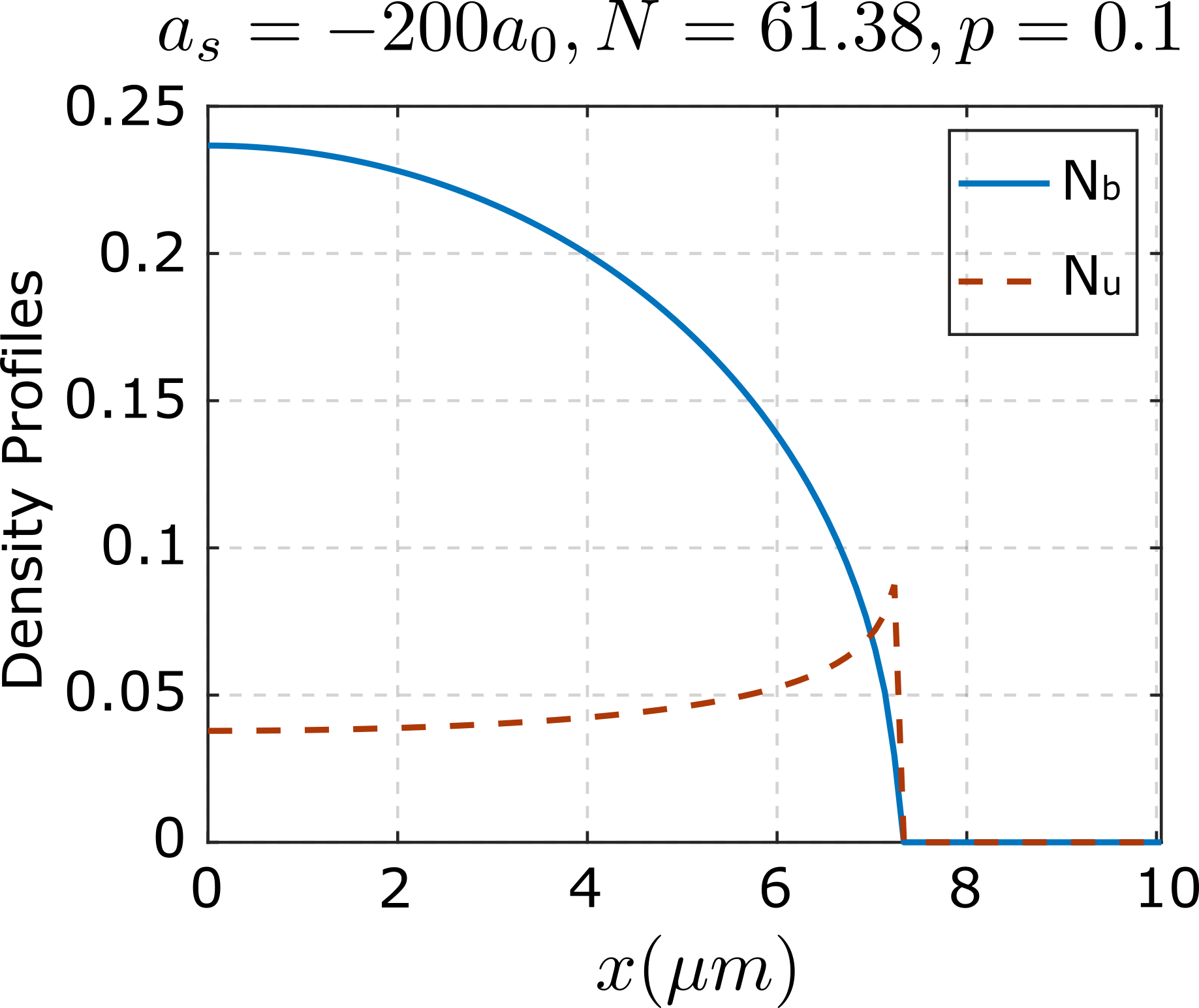}}
		\subfigure[]{\includegraphics[scale=0.68]{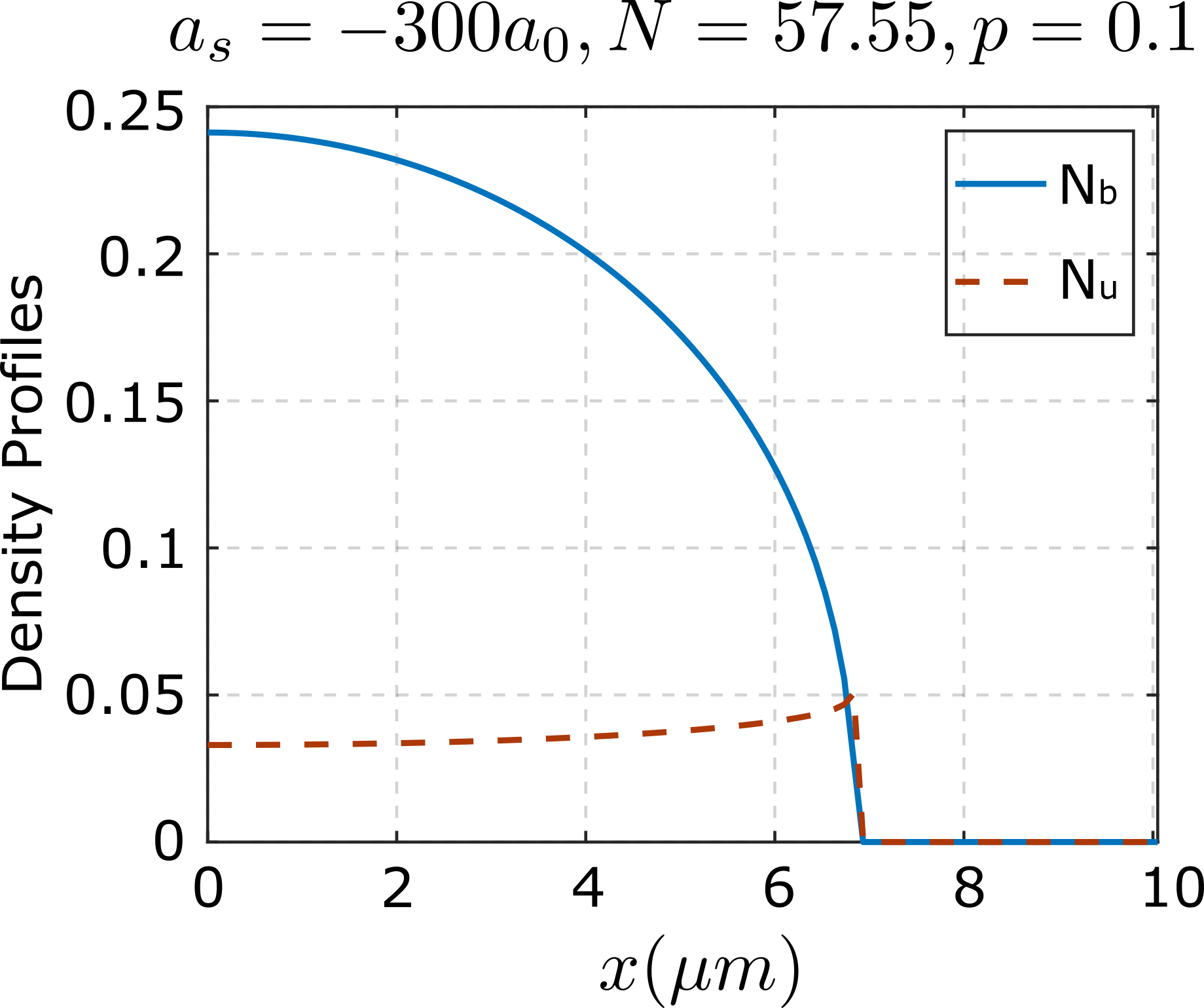}}
		\subfigure[]{\includegraphics[scale=0.68]{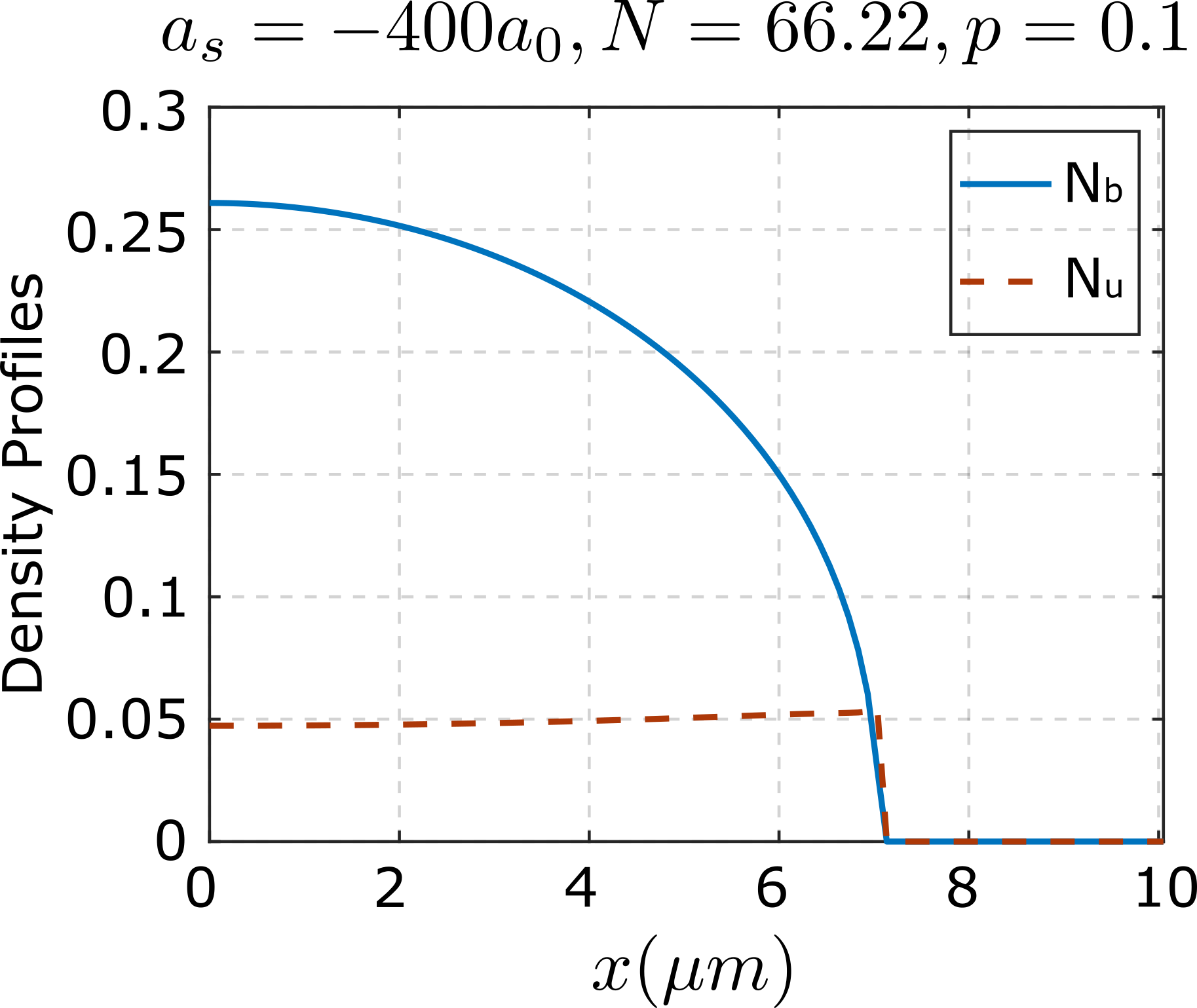}}
		\subfigure[]{\includegraphics[scale=0.68]{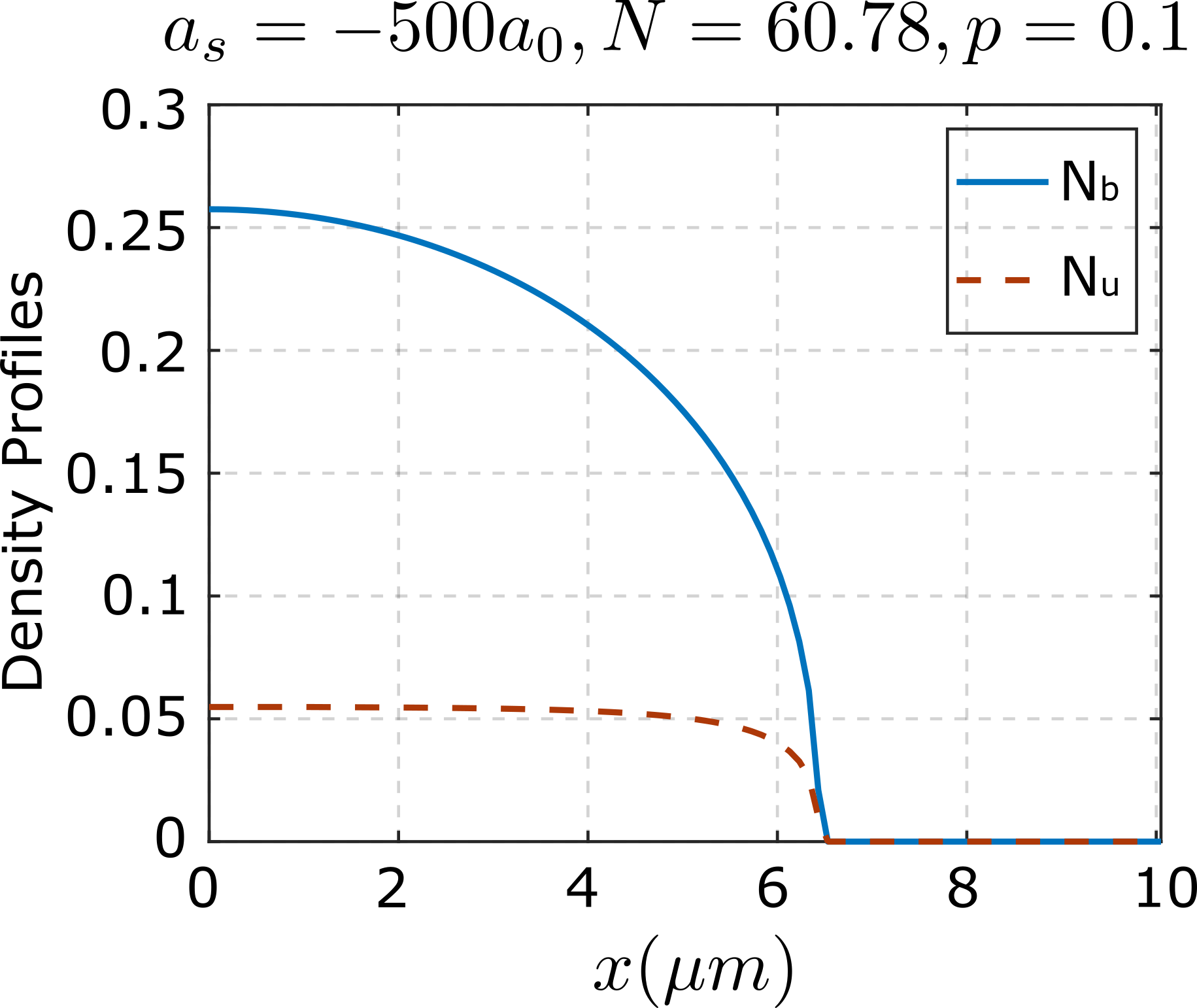}}
        \subfigure[]{\includegraphics[scale=0.68]{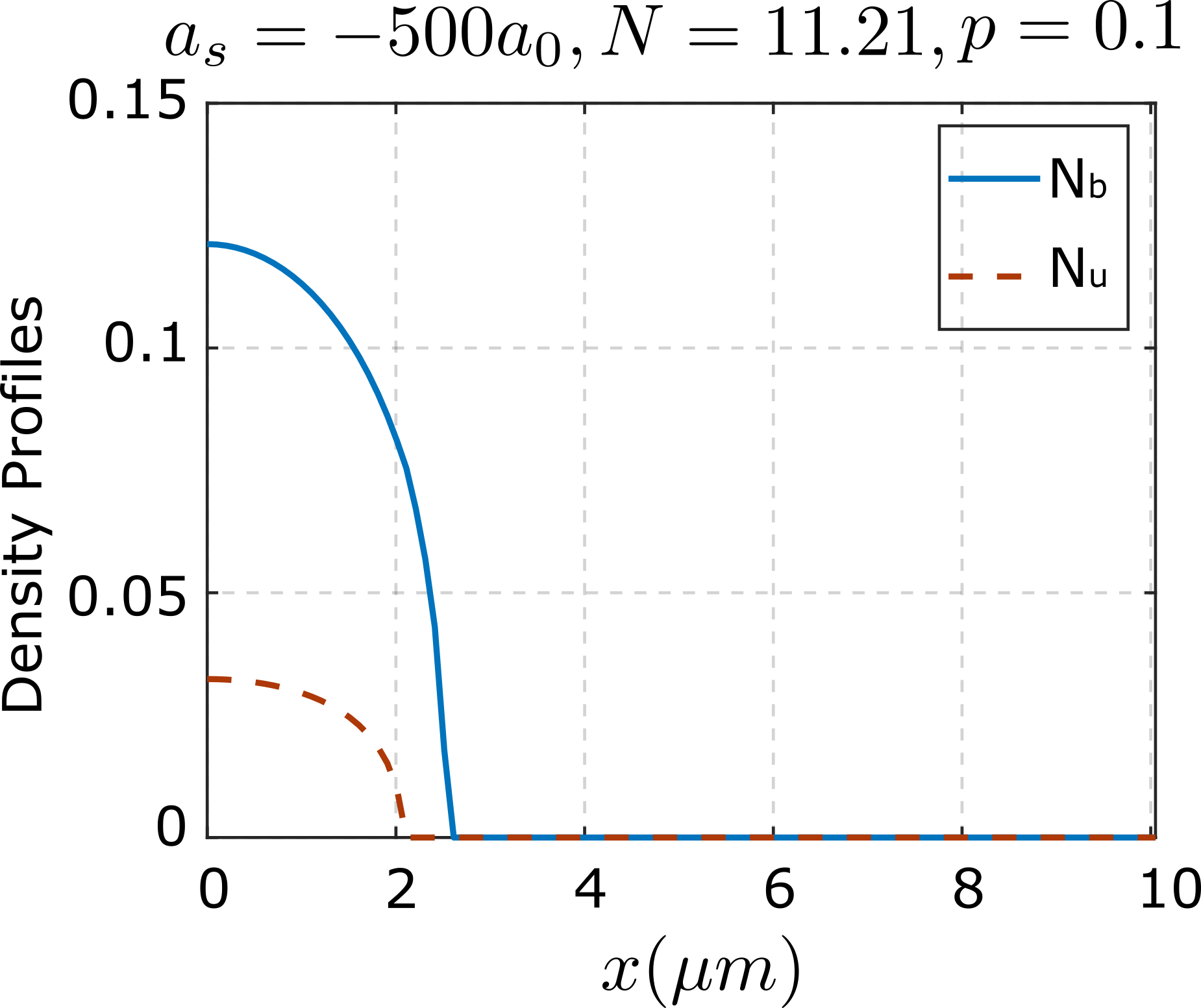}}
		\caption{The density profiles of the paired fermions ($N_b$, blue solid lines) and unpaired fermions ($N_u$, red dashed lines) in a tube with the experimental settings, but at zero temperature. For large value of $N \sim 60$, the radius of the unpaired fermions extends to a very large region of the tube with $p$ larger than or close to $p_c$, see (a)-(e). This leads to a spin gapless excitation which dominates the low-energy excitations in the mixed phase. Whereas, for a tube with small $N$, once again the trapped gas exists in two phases, i.e., a mixed phase in the trap center and a pure paired phase in the two wings, see (f).} 
		\label{Fig1}
	\end{figure}

In Fig.~\ref{Fig1}, we show the density profiles in a single tube with polarization $p=0.1$, which is close to the critical polarization $p_c$, predicted theoretically \cite{Orso:2007} and confirmed experimentally \cite{Liao:2010}. For $p<p_c$, a harmonically confined 1D attractive Fermi gas will phase separate into a mixed phase, located in the center of the tube, with fully paired phases at either end, while for $p>p_c$, the end regions are fully polarized. 
The value of $N$ for the tube was obtained from the experimental data by averaging approximately 20 shots for each set of parameters. The calculated DSFs used the average value of $N$ for each panel. We find that small $p$ explains the apparent gapless excitations in the spin sector. 
If we reduce $p$ or $N$, the central region is a mixed phase consisting of both paired and unpaired fermions, while the two wings are fully paired, the size of which increases as the magnitude of the interaction strength increases. In order to make the full paired region dominate the whole atomic cloud, it is necessary to make $p$ very close to zero.

 
	 %

	 \section{III. Spin and Charge Dynamical Structure Factor}

     Bragg spectroscopy measures the dynamical structure factor (DSF). In a homogeneous 1D tube with both paired and unpaired fermions,
the charge DSF reads
	   \begin{eqnarray}\label{27}
	   	S_c(q,\omega)=S_u(q,\omega)+S_b(q,\omega)
	   \end{eqnarray}
	   where, in terms of the retarded density Green function
\[ \chi_{u(b)}(q,\omega)=-\mathrm{i} \theta(t)\int dx \int dt e^{-\mathrm{i}(q x-\omega t)} \langle [n_{u(b)}(x,t) ,n_{u(b)}(0,0)]\rangle ,\]
the DSFs for unpaired and paired fermions are given by
	    \begin{eqnarray}\label{28}
	   	S_{\alpha}(q, \omega)=\operatorname{Im} \chi_{\alpha}(q, \omega)
	   	=\frac{N  }{2 \hbar q v_{\alpha}} \!\left\{ \frac{1}{\exp\left[ \beta \frac{m^*_\alpha}{2}(v^2_- -v^2_\alpha) \right]+1}-\frac{1}{\exp\left[ \beta \frac{m^*_\alpha}{2}(v^2_+-v^2_\alpha) \right]+1}\right\},
	   \end{eqnarray}
	   where $\alpha= u,\,b$; $v_\pm \equiv \frac{\omega}{q} \pm \frac{\hbar q}{2m^*_\alpha}$, $m^{*}_{\alpha}$ and $v_{\alpha}$ are the effective mass and sound velocity of the unpaired (paired) fermions, respectively, which can be obtained from Bethe Ansatz. Here, we have approximated paired and unpaired fermions as two independent non-interacting Fermi gases with Fermi velocity given by $v_\alpha$ and mass $m^*_\alpha$.
	 
	   The spin DSF $S_s(q,\omega)$ is the imaginary part of the retarded spin density Green function, $S_s=\operatorname{Im}\left[\chi_s(q,\omega)\right]$,
	   where
	  \begin{eqnarray}
	  \chi_{s}(q,\omega)=-\mathrm{i} \theta(t)\int dx \int dt e^{-\mathrm{i}(q x-\omega t)} \langle [S^z(x,t) ,S^z(0,0)]\rangle.\label{17}
	  \end{eqnarray}
	   The spin density can be expressed as  
	  \begin{eqnarray}
	  	S^z(x,t)=\frac{1}{2}\left[n_{\uparrow}(x,t)-n_{\downarrow}(x,t)\right] \approx \frac{n_{u}(x,t)}{2}\label{18}
	  \end{eqnarray}
	 Therefore, according to the asymmetric spin detuning in performing spin Bragg spectroscopy, the bound pairs are insensitive to the response. Whereas, the spin DSF is actually equivalent to the DSF of unpaired fermions, i.e.,
	  \begin{eqnarray} 	 S_s(q,\omega)=S_u(q,\omega)={\rm Im}\left[\chi_u(q,\omega)\right].\label{19}
	  \end{eqnarray}

	   \section{IV. Spin DSF in the fully paired phase}
	In the mixed phase with both paired and unpaired fermions, the spin DSF is dominated by the latter. Here, we first consider what happens in a fully paired system under the framework of the Tomonaga-Luttinger liquid (TLL) theory.  
    
	   The correlation function of a gapped system will acquire an exponentially decaying factor $\sim e^{-\Delta\sqrt{\tau^{2}+v^{2}x^{2}}}$ \cite{2}. Thus, the spin density $s^z(x,\tau)$ correlation function is approximately:
	   \begin{eqnarray}
	   	\chi^{\Delta}(x,\tau)=\left\langle(s^z(x,\tau))s^z(0,0)\right\rangle\sim \chi^{TLL}(x,t)e^{-\Delta\sqrt{(\tau)^{2}+v^{2}x^{2}}},\label{22}
	   \end{eqnarray}
	   where $\chi^{TLL}$ is the spin density correlation function of the gapless system and can be calculated using the TLL theory as
	   \begin{eqnarray}
	   	\chi^{TLL}_{q\sim0}(x,\tau)&=&\frac{K}{2\pi^{4}}\left[\frac{1}{\left(u\tau+ix\right)^{2}}+\frac{1}{\left(u\tau-ix\right)^{2}}\right]\label{23}
	   \end{eqnarray}
       where $K$ is the Luttinger parameter and $u$ the spin velocity. 
	   After a Fourier transformation and analytic continuation of $\chi^{TLL}$, we get the retarded correlation function for the gapless system as
	   \begin{eqnarray}
	   	\chi(q,\omega)|_{i\omega\to \omega+i\delta}=\frac{K}{\pi^{3}}\frac{-2uq^{2}}{(\omega+i\delta)^{2}-(uq)^{2}}\label{24}
	   \end{eqnarray}
	   whose imaginary part is the DSF
	   \begin{eqnarray}\label{25}
	   	S(q,\omega)={\rm Im}\left[\chi(q,\omega)\right]=\frac{Kq}{\pi^{2}}\left[\delta(\omega-uq)-\delta(\omega+uq)\right],
	   \end{eqnarray}
	   and the peak frequency of the DSF is determined by $uq$.
	   
	   For the gapped system, we need to consider the Fourier transformation of $\chi^{\Delta}$ to obtain the DSF. This is generally very complicated. However, for a large gap $\Delta$, the correlation length is short, and we obtain the following approximate form: 
	   \begin{eqnarray}\label{26}
	   	S^{\Delta}(q,\omega)=\frac{Kq}{\pi^{2}}\left[\delta(\omega-(uq+\Delta))-\delta(\omega+(uq+\Delta))\right].
	   \end{eqnarray}
	   Comparing Eq.~\eqref{25} and Eq.~\eqref{26}, we can see that, in contrast to the gapless system, the peak frequency of the spin DSF of the gapped one should be shifted to a higher frequency by an amount $\Delta$.

	  \section{V. Fitting with the experiment data}

\begin{figure}[H]
	 	\centering
	 	\includegraphics[scale=1]{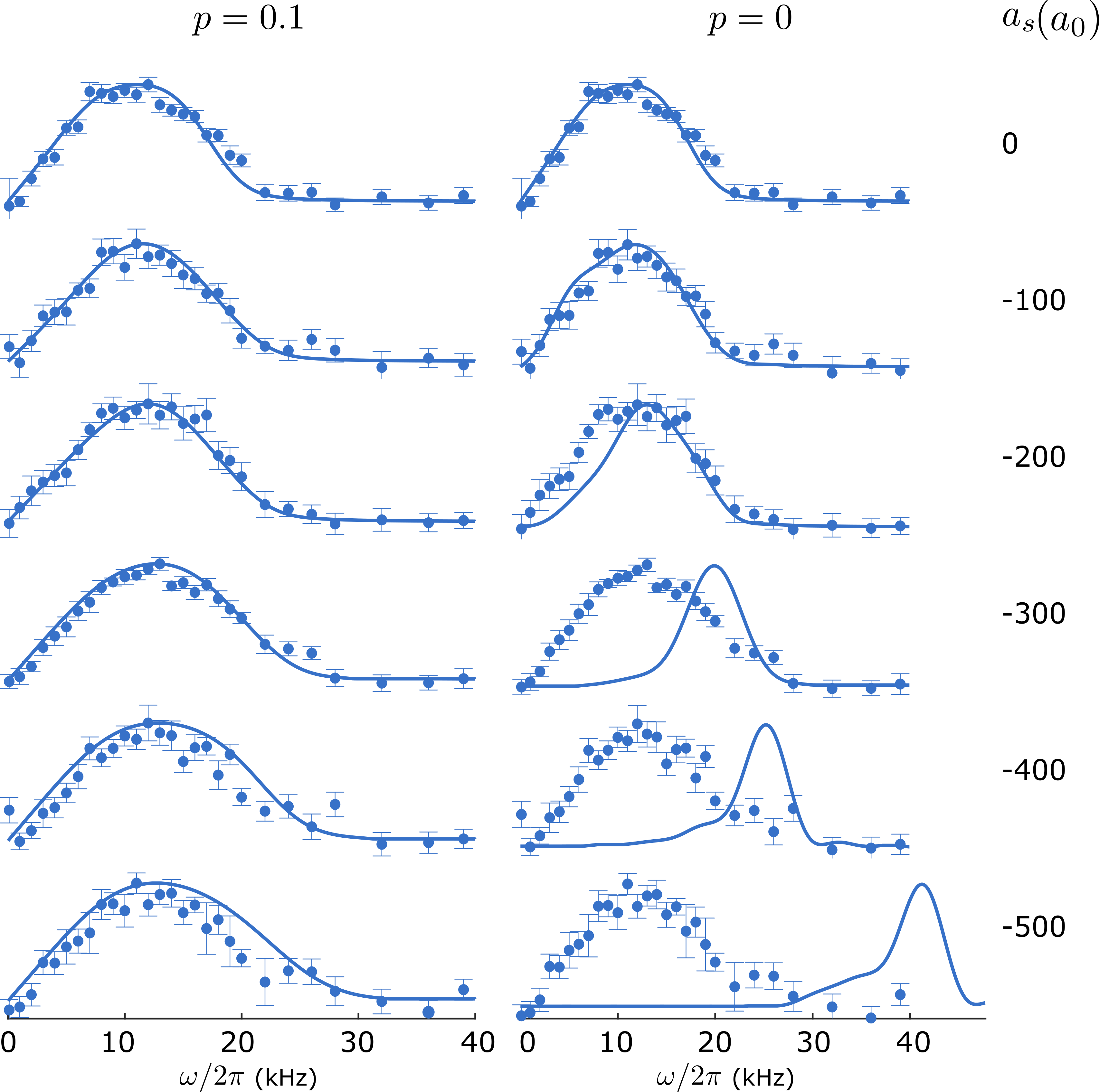}
	 	\caption{Spin DSF experimental (discrete points) and calculated (solid lines). In the left column, the calculated DSF's assume $p=0.1$, which gives the best fit to the experimental data for all scattering lengths shown. The right column corresponds to the spin-balanced case with $p=0$ for all scattering lengths shown. }
	 	\label{Fig2}
	 \end{figure}

\begin{figure}[H]
	 	\centering
	 	\includegraphics[scale=1]{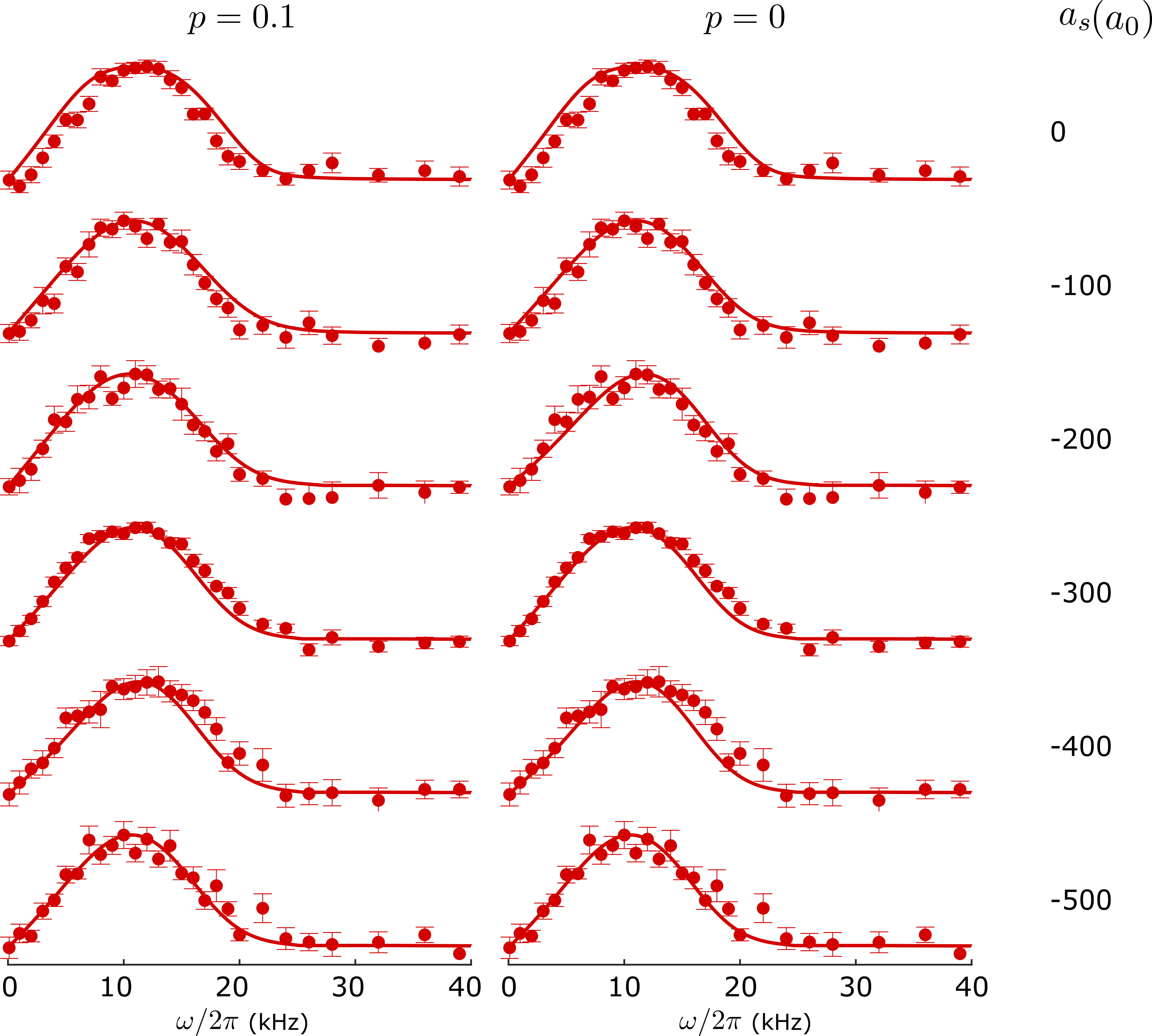}
	 	\caption{Same as Fig.~\ref{Fig2} except for the charge mode.}
	 	\label{Fig3}
	 \end{figure}

We carefully simulate the experimental data using the above theoretical formula with parameters that are the same as the experimental setting. 
	 We assume that there is a small polarization ($p=0.1$) in each tube, and fit to the spin and charge DSFs as shown in the left panels of Fig.~\ref{Fig2} and Fig.~\ref{Fig3}, respectively. 
	 We find that the theoretical results fit well to the experimental data under the assumption of a small polarization.
	  Based on the theoretical results presented above, we find that the spin Bragg signal is dominated by the contribution from the mixed phase in the center of the tubes, showing the signature of a spin gapless excitation. The gapped excitation from the fully paired phase in the outer wings has a negligible contribution to the spin DSF. 

      In the right panels of Fig.~\ref{Fig2} and Fig.~\ref{Fig3}, we show the spin and charge DSF of a spin-balanced system ($p=0$), respectively. Comparing the left and right panels, we immediately see that whether or not a small polarization is present does not qualitatively affect the charge DSF. This is because the low-energy charge excitation is gapless, regardless of whether a spin imbalance is present or not. In contrast, Fig.~\ref{Fig2} shows that a spin imbalance plays an essential role for the spin DSF. In particular, for strong attraction ($|a_s|>200 a_0$), the entire system is fully paired with a gapped spin excitation when $p=0$. A clear spin gap emerges in the spin DSF (see right panel of Fig.~\ref{Fig2}). In comparison, spin excitation is always gapless in the presence of spin imbalance.

\begin{figure}[H]
	 	\centering
	 	\includegraphics[scale=1.0]{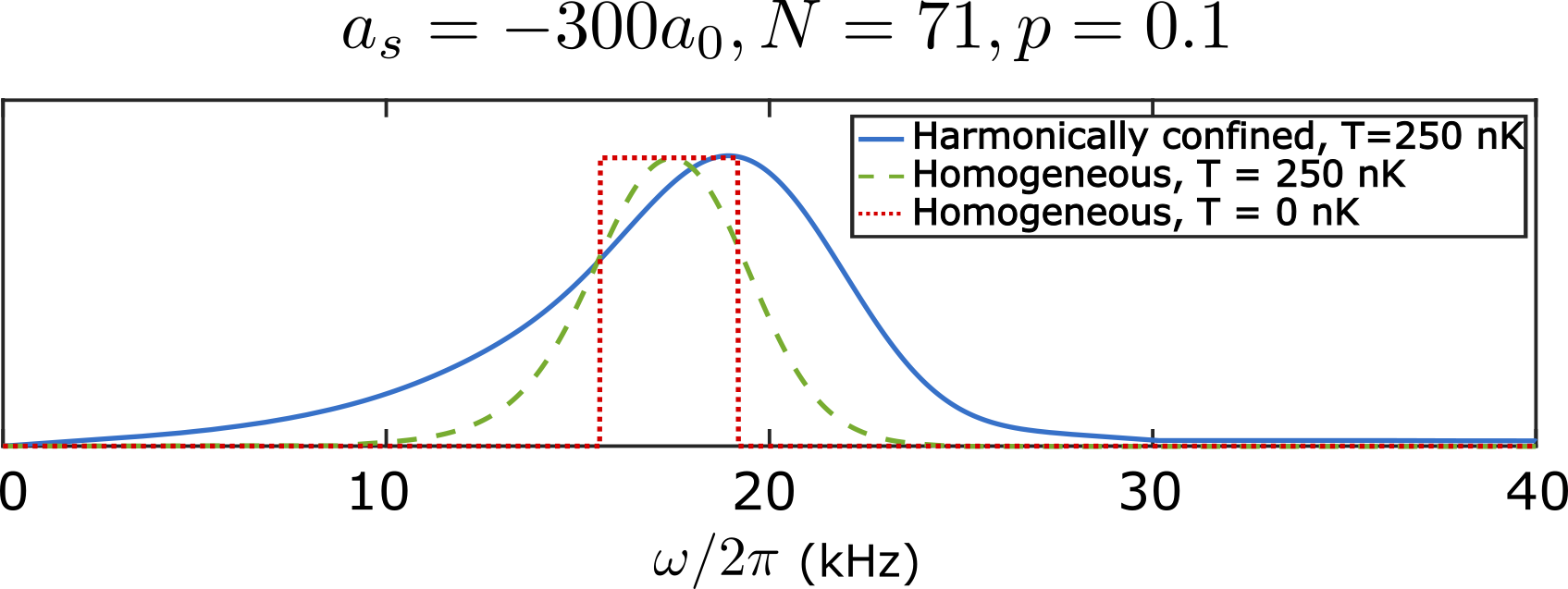}
	 	\caption{Within the approximation of the previous sections, the calculated spin Bragg spectrum of a single tube with $a_s=-300a_0$, $N=71$, $p=0.1$. The solid blue line is obtained by summing the DSFs of all cells at $T=250$ nK. Whereas, for a homogeneous gas (assuming the density of which is the same as the density in the trap center for the trapped case), the spin DSF presents a rectangular shape (dotted red line) at $T=0$, and a peak (dashed green line) at $250$ nK. }
	 	\label{Fig4}
	 \end{figure}

	    In Fig. \ref{Fig4}, we show the spin Bragg spectrum of a single tube for a small polarization ($p=0.1$). We see that the smooth peak is a consequence of low temperatures (250 nK) and averages over the inhomogeneous density profile. For a homogeneous gas, the spin DSF has a rectangular shape at zero temperature,  whereas it shows a sharp peak at low temperatures. 
	  If the number of particles in the tube increases, the peak moves further to the right (i.e., higher energy). On the other hand, averaging the DSFs of all tubes with smaller numbers of particles makes the peak even rounder and to move further to the left. The gapless nature of the low-energy excitations in both the spin and charge sectors explains the similarity of the DSF's.


